\documentclass[aps,preprint,superscriptaddress]{revtex4-1}

\usepackage{graphicx}
\usepackage{epstopdf}
\usepackage{amsmath}

\begin{document}
\title{From jamming to collective cell migration through a boundary induced transition}

\author{Oleksandr Chepizhko}
\affiliation{Institut f\"ur Theoretische Physik, Leopold-Franzens-Universit\"at Innsbruck, Technikerstrasse 
21a, A-6020 Innsbruck, Austria}
\author{Maria Chiara Lionetti}
\affiliation{Center for Complexity and Biosystems,
Department of Environmental Science and Policy,  University of Milan, via Celoria 26, 20133 Milano, Italy}
\author{Chiara Malinverno} 
\affiliation{IFOM, the FIRC Institute of Molecular Oncology, Via Adamello 16, 20139 Milano, Italy}
\affiliation{Dipartimento di Scienze della Salute, San Paolo, University of Milan,
		20122 Milano, Italy}
	\author{Giorgio Scita}
	\affiliation{IFOM, the FIRC Institute of Molecular Oncology, Via Adamello 16, 20139 Milano, Italy}
	\affiliation{Dipartimento di Scienze della Salute, San Paolo, University of Milan,
		20122 Milano, Italy}
\author{Stefano Zapperi} 
\affiliation{Center for Complexity and Biosystems,
Department of Physics, University of Milano, via Celoria 26, 20133 Milano, Italy}
\affiliation{CNR - Consiglio Nazionale delle Ricerche,  Istituto per l'Energetica e le Interfasi, Via R. Cozzi 53, 20125 Milano, Italy}
\affiliation{Department of Applied Physics, 
Aalto University, P. O. Box 11100, FIN-00076 Aalto, Espoo, Finland}
\author{Caterina A. M. La Porta}\affiliation{Center for Complexity and Biosystems,
Department of Environmental Science and Policy,  University of Milan, via Celoria 26, 20133 Milano, Italy}
\email{\rm Corresponding authors: stefano.zapperi@unimi.it;caterina.laporta@unimi.it}

\begin{abstract}
Cell monolayers provide an interesting example of active matter, exhibiting a 
phase transition from a flowing to jammed state as they age. Here we report experiments and numerical simulations illustrating how a jammed cellular layer rapidly reverts to a flowing state after a wound. Quantitative comparison between experiments and simulations shows that cells  change their self-propulsion and alignement strength so that the system crosses a
phase transition line, which we characterize by finite-size scaling in
an active particle model. This wound-induced unjamming transition 
is found to occur generically in epithelial, endothelial and cancer cells. 
\end{abstract}
 \maketitle

\section{Introduction} 

Understanding collective cell migration, when cells move as a cohesive and coordinated group is important  to shed light on key aspects of embryogenesis, wound repair and cancer metastasis \cite{Friedl2009}. While cellular and multicellular dynamics and motility is controlled by a complex network of biochemical pathways \cite{Ilina2009}, it is becoming increasingly clear that a crucial role is also played by physical interactions among cells and between cells and their environment\cite{Tambe2011,Brugues2014,Haeger2014,Lange2013,koch2012}. Dense cellular assemblies, such as epithelial monolayers or cancer cell colonies, share many features with amorphous glassy materials, displaying slow relaxation\cite{Angelini2011}, jamming \cite{Park2015} and intermittent avalanche fluctuations \cite{Chepizhko2016}. 

Cells can collectively flow like a fluid, but  as cell proliferation increases cell density mutual crowding leads to slowing down and arrest \cite{Angelini2011,Park2015}. The phenomenon of cell jamming could have a functional biological role, ensuring the development of tissue elasticity and protective barriers in epithelial tissues, providing also a suppressive mechanism for the aberrant growth of oncogenic clones. An unjamming transition into a collective flowing state could  also provide a mechanism for cell migration that is alternative to the well studied epithelial-to-mesenchymal transition (EMT), characterized by individual cell migration as a result of the loss of intracellular adhesion \cite{ye2015}.

The loss of motility in cellular assemblies shares similar features to the jamming transition observed in disordered materials such as colloids, granular media or foams \cite{Liu2010}. As in disordered solids, cell jamming is thought to occur across different routes, either through the increase of cell density, or the reduction of the active forces responsible for cell motility. In analogy with inanimate glassy systems, is seems reasonable to assume that jamming could also be produced by increasing intracellular adhesion, yet experimental observations show that unjamming is associated with an increased adhesion, questioning the role of ashesion as a principal determinant of jamming \cite{Park2015}. Recent experiments also show that over-expression of the endocytic master regulator RAB5A leads to rapid fluidization and 
unjamming of a jammed confluent cell monolayer, due to polarization of cell protrusion and increase 
in traction force \cite{Malinverno2017}.

Theoretical understanding of the cell jamming transition mostly relies on simulations of
vertex models \cite{Bi2015}, similar to those used for foam rheology \cite{Weaire1984,Okuzono1995}, 
and self-propelled Voronoi (SPV) models \cite{Li2014,Bi2016}, also including self-propulsion typical
of active particle models\cite{Szabo2006,Poujade2007,Sepulveda2013,Vedula2013}. In the SPV model, space is subdivided into Voronoi polygons endowed with a mechanical energy due to cell surface tension, compression and adhesion. The polygonal structure then evolves according to equation of motion taking into account an active self-propulsion force and a noise term. Extensive numerical simulations allow us to identify the jamming transition and to reconstruct a possible phase diagram in terms of the main physical parameters encoded in the model \cite{Bi2016}. In this way, it is possible to identify  physical determinants for jamming, including  the strength of the self-propulsion force and the shape anisotropy of the cell, determined by the ratio between the cell area and its squared perimeter \cite{Park2015,Bi2016}. 

While cell jamming is typically observed experimentally by following in time the evolution of a cell monolayer in a confined space, collective cell migration is mostly observed and quantified in wound healing assays  where cells are allowed to invade an empty space \cite{Tambe2011,Brugues2014,Haeger2014,Lange2013,koch2012,Poujade2007,Sepulveda2013,Vedula2013}. The implementation of current vertex and Voronoi models rely on periodic boundary conditions and are thus not appropriate to study wound healing, whose statistical features are well described by active particle models\cite{Szabo2006,Sepulveda2013,Chepizhko2016}.  
In this paper, we analyze the response of a  jammed cell monolayer to the appearance of a wound, which creates empty space for the cells at the boundary. To this end, we analyze a large set of time-lapse images of cell layers using different cell types and experimental treatment, under confluent and wound healing conditions, and compare the results with simulations of an active particle model. We show that the model is able to quantitatively describe  experimentally measured velocity distributions with open but also with confined boundary conditions. The model allows for a systematic investigation of the role of each relevant  physical parameter so that we can place each experiment into a suitable phase diagram. Our simulations show that the model exhibits a clear jamming transition characterized by finite-size scaling. When we compare wound healing with confined experiments, we find that the latter are described by a set of physical parameters that place the system into the jamming phase, while the former fall into the flowing phase. Hence, the presence of a wound re-awakens jammed cells in a way similar to the effect of the  overexpression of the endocytic regulator RAB5A \cite{Malinverno2017}. 

\section{Experimental results}
We consider time lapse images recorded during wound healing assays performed for a variety of cell lines under different experimental conditions (see section \ref{sec:methods} for detailed information): HeLa cells on different substrates: plastic, soluble collagen, or fibrillar collagen (see \cite{Chepizhko2016}); Endothelial cells derived from embryonic stem cells with homozygous null
mutation of the VE-cadherin gene (VEC null) \cite{balconi2000} and those with the corresponding
wild type form of VE-cadherin (VEC positive) \cite{Lampugnani2002}; Endothelial cells isolated from lungs of wild type adult mice (lung ECs) \cite{giampietro2015,Chepizhko2016}; A non-tumorigenic human mammary epithelial cell line (MCF-10A) with the over-expression of RAB5A, known to induce unjamming, and the corresponding  wild type (WT) cells \cite{Malinverno2017}. As a comparison, we also analyze HeLa cells and and MCF-10A cells in confluent conditions without inducing wound healing.

\begin{figure*}[htb]\centering 
\includegraphics[width=16cm]{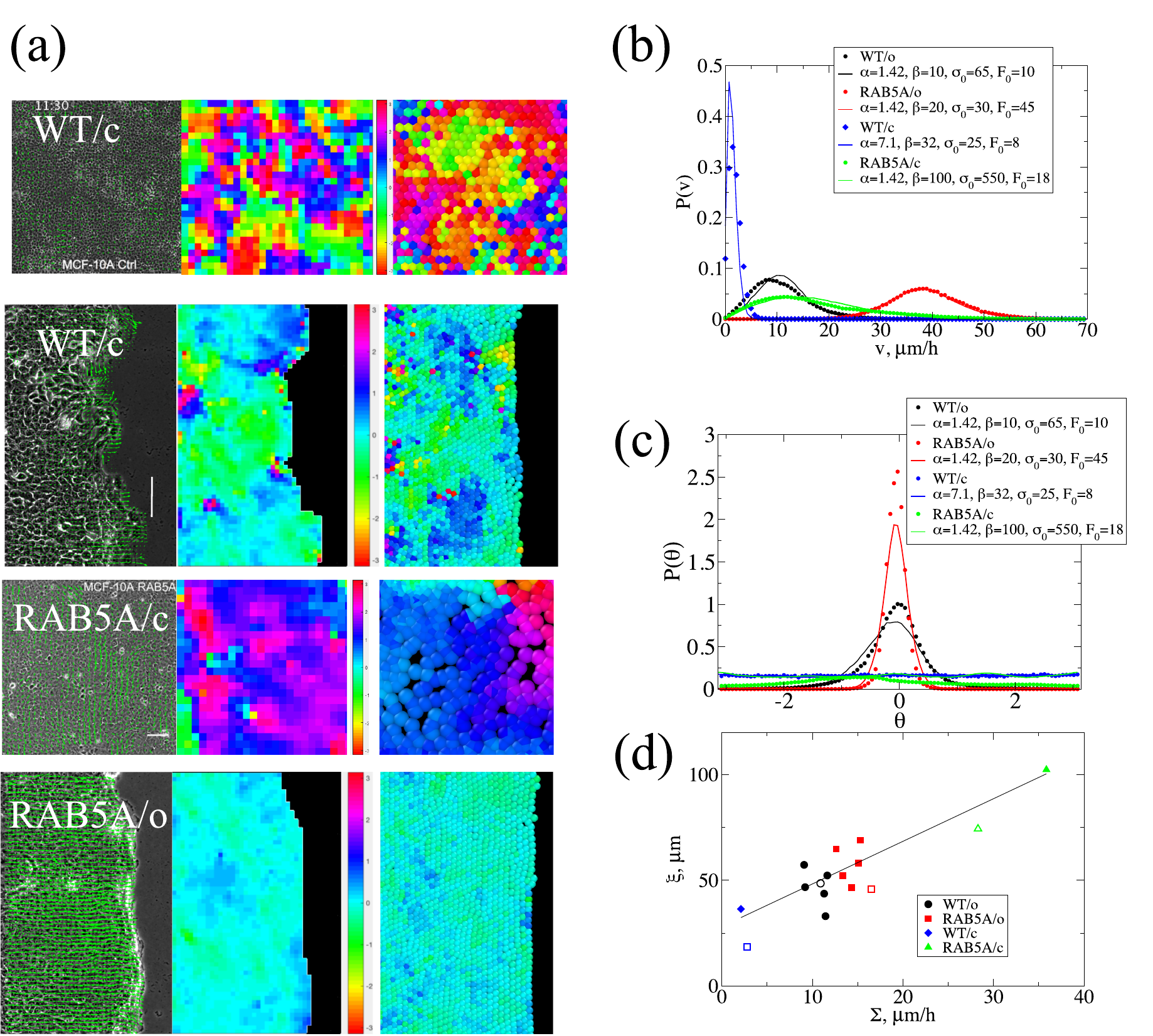}
 \caption{\label{fig:1} Statistical properties of collective migration in epithelial cells. a)  Velocity maps obtained from PIV for  WT MCF-10A cells in wound healing (WT/o) and confluent (WT/c) conditions, as well as MCF-10A cells with RAB5A overexpression under the same conditions (RAB5A/o and RAB5A/c, respectively). Arrows represent the velocities and the color their orientations. Experimental results are compared with simulations.
b) The distribution of the absolute values of the velocities for the experiments reported in panel a) compared with the corresponding simulation results. c) Distribution of velocity
orientations for the same cases. d) Velocity correlation length $\xi$ as a function of the velocity fluctuations $\Sigma$. The linear regression was predicted by an earlier model \cite{Garcia2015}.}
 \end{figure*}

We analyze all the experimental time lapse images using particle image velocimetry (PIV) 
which allows us to estimate the local velocity field inside the cell sheet (See Fig. \ref{fig:1}a)
and the corresponding velocity distributions  \cite{Chepizhko2016}.
In particular, we measure the distribution of the velocity absolute values $P(v)$  (See Fig. \ref{fig:1}b), the distribution of the velocity orientations (See Fig. \ref{fig:1}c) and the velocity-velocity correlation
lengths $\xi$ from an exponential fit of the correlation functions (See Fig. \ref{fig:1}c). It is
interesting to notice that correlations lengths obtained from MCF-10A cells under different conditions
scale linearly with the standard deviation of the velocity distribution
$\Sigma = \sqrt{\langle (v_x^2+v_y^2)\rangle - (\langle v_x \rangle)^2-(\langle v_y \rangle)^2)}$.
This relation was predicted by a theoretical model of collective cell migration where it implies
that the effective friction is dominated by cell-cell interactions over cell-substrate interactions
\citep{Garcia2015}.

Considering all the experimental results, we obtain a set of 40 distributions for different
cell lines and experimental conditions describing the statistical features of the velocity
during collective cell migration. To visualize all the data in a simple way, in Fig. \ref{fig:2}a 
we report the average velocity $\langle v_x \rangle$ perpendicular to the front as
a function of $\Sigma$.
The results show that the standard deviation of confluent systems is roughly 
one order of magnitude smaller than the one observed during wound healing for the same cells. 
Considering the wound healing case, we observe that MCF-10A cells move considerably faster
than the other cells and, as expected, RAB5A overexpression leads to net increase in the 
average velocity with little changes in $\Sigma$.

\section{Model}
In order to relate the statistical properties of collective cell migration quantified
from our experiments with the presence of a jamming critical point, we compare the
expperimental results with simulations of a model for interacting cells similar to the one introduced in Ref. \citep{Sepulveda2013}.  Cells are treated as particles moving according
to the following equation of motion:
\begin{equation}
  \label{eq:main_Sepulveda_simple}
  \frac{d \mathbf{v}_i}{d t} =
        - \alpha \mathbf{v}_i + \sum_{j n.n. of i}
        \left[
          \frac{\beta}{N_i}(\mathbf{v}_j-\mathbf{v}_i)+\mathbf{f}_{ij}
          \right]
        +
        \sigma \boldsymbol{\eta}_i + 
        \mathbf{\hat{v}}F_0\,.
\end{equation}
The first term represents the dissipation processes and $\alpha$ is a damping parameter. The 
second term represents the interaction of $i$-th cell with the other cells. The interaction reflects the tendency of a particle to orient its velocity with the velocity of its neighbors,
with a  coupling strength $\beta$, and $N_i$ is the number of neighbors of $i$-th cell. The second interaction term, $\mathbf{f}_{ij}$, represents the short-range hard-core repulsion and long-range attraction, and has the form
\begin{equation}
  \label{eq:fij}
  \mathbf{f}_{ij}=-\nabla_i U(r_{ij})\,,
\end{equation}
where
\begin{equation}
  \label{eq:potential}
  U(r)=U_0 \exp(-(r/a_0)^2)+U_1 (r-a_1)^2 H(r-a_1)\,,
\end{equation}
where $H(x)$ is the Heaviside function $H(x)=1$ for $x>0$ and $H(x)=0$ otherwise.
The motion of a cell is affected by a noise term $\sigma_0 \boldsymbol{\eta}_i$, where $\boldsymbol{\eta}_i$ is an Ornstein-Uhlenbeck process with correlation time $\tau$:
\begin{equation}
  \label{eq:OUprocess}
  \tau \frac{d \boldsymbol{\eta}_i}{dt} = - \boldsymbol{\eta}_i + \boldsymbol{\xi}_i,
\end{equation}
$\mathbf{\xi}_i$ is a delta-correlated white noise, independent for each cell $\langle \mathbf{\xi}_i (t) \mathbf{\xi}_j (t') \rangle = \delta_{ij} \delta(t-t')$.

As in the original model \cite{Sepulveda2013}, the free surface is modeled by surface particles, that are hindering cells invade the space. The interaction between a surface particle and a cell is modeled by:
\begin{equation}
  \label{eq:fs}
  \mathbf{f}_{ij}^s=-\nabla_i U^s(r_{ij})\,,
\end{equation}
where $U^s(r)$ is $U^s(r)=A_s \exp(-(r/a_s)^2)$.
With each surface particle a scalar damage variable $q$ is associated, and it is chosen to obey:
\begin{equation}
  \label{eq:dtheta}
  \nu \frac{d q_i}{dt}=\sum_{j: \mathbf{r_j} \in S_i} |\mathbf{f_{ij}^s}|\,.
\end{equation}
The surface particle disappears if the damage variable reaches its critical value $q_i=q_c$.
To simulate wound healing conditions, we place the cell in a box of size $2L \times L$
in which we place cells surrounded by surface particles. We first do not allow damage on
the surface particles and let the system relax for a time $T/4$. We then
allow surface particles to be damaged so that the wound healing process starts. We first let
the front evolve for an additional time $T/4$ and finally collect velocity statistics
for a time $T/2$. This procedures allows to minimize transient effects.
The model can also be simulated in confluent conditions by setting the damage variable
to $q_c=\infty$, so that the cells never invade the surrounding space. 

\section{Simulations}

\begin{figure}[htb]\centering 
\includegraphics[width=\columnwidth]{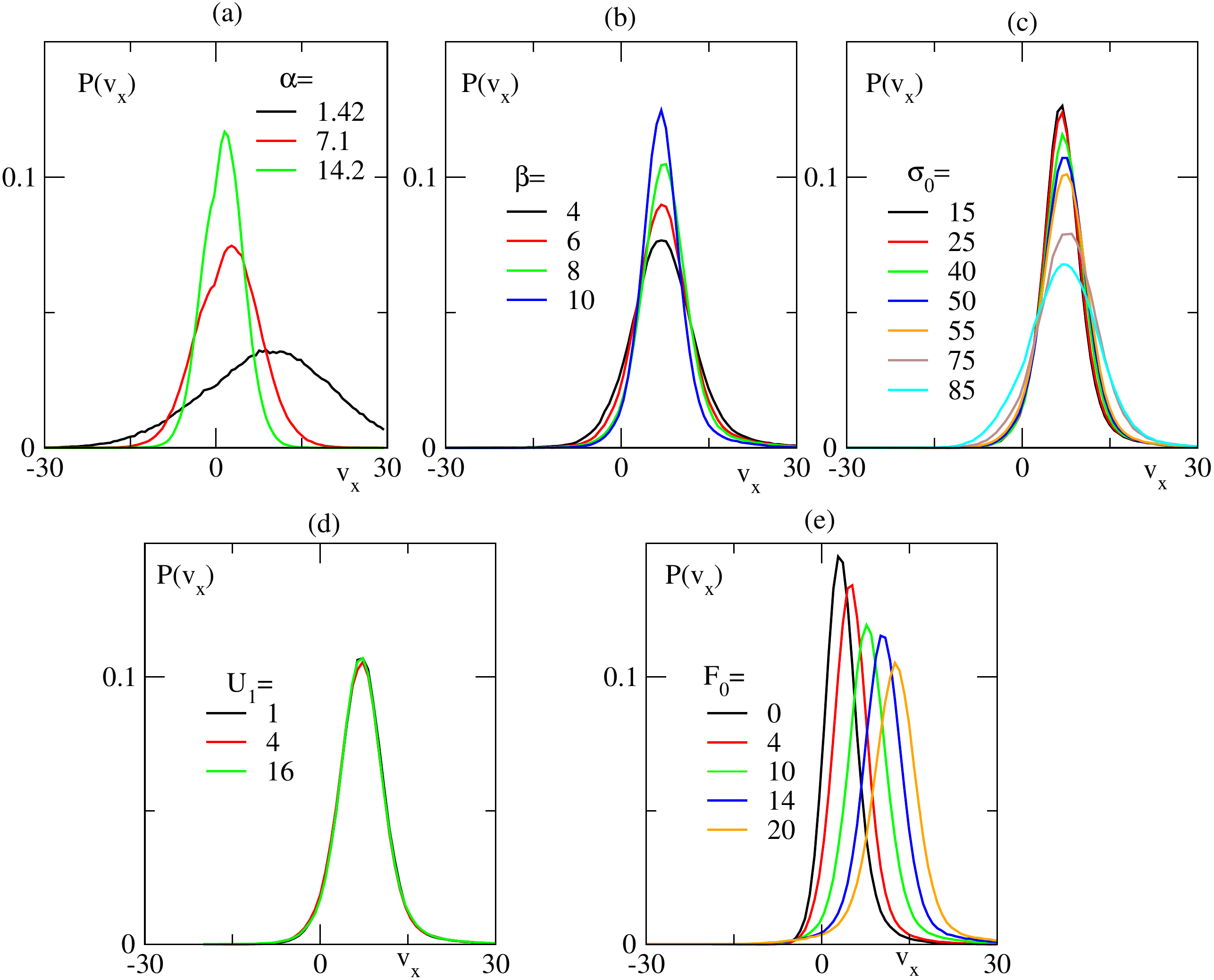}
 \caption{\label{fig:S1} Effect of the model parameters on the cell velocity distribution. Effect of
 a) damping $\alpha$, b) alignment strength $\beta$, c) noise $\sigma_0$, d) adhesion $U_0$ and 
 e) self-propulsion force $F_0$ on $P(v_x)$, where $v_x$ is the velocity in the invasion direction
 in wound healing conditions.}
 \end{figure}

\subsection{Experimental velocity distribution are described by the model}
To compare the experiment with the model, we run a large set of simulations with both open
and closed boundary conditions, employing a cell density that corresponds to the one
observed in experiments for confluent cell layers. We explore the parameter space defined by different values of ($F_0$, $\sigma_0$, $\beta$, $\alpha$, $U_1$) and inspect how 
each parameter affects the velocity distributions (see Fig. \ref{fig:S1}). 
We observe that $F_0$ and $\alpha$ have a considerable effect on $\langle v_x \rangle$, while $\beta$ and
$\sigma_0$ have opposite effects on the velocity fluctuations: $\Sigma$ increases with $\sigma_0$ and decreases with $\beta$. The adhesion strength is found to have little effect on the velocity distributions for the range we examined. 

\begin{figure}[htb]\centering 
\includegraphics[width=\columnwidth]{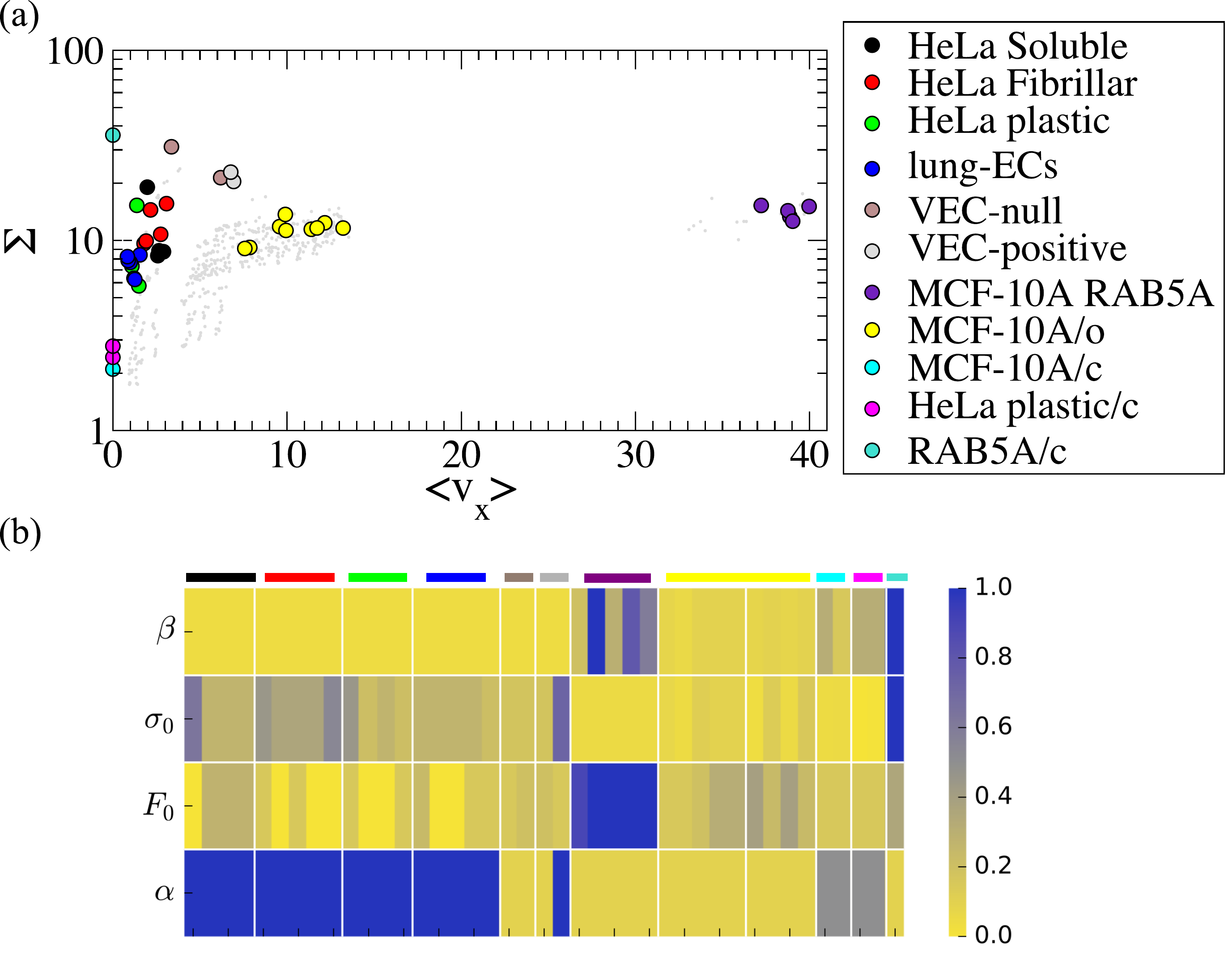}
 \caption{\label{fig:2} The parameter space of cell migration.
 a) A map of the average velocity in the invasion direction
$\langle v_x \rangle$ and the velocity standard deviation $\Sigma$ for all the cell lines
analyzed. b) A colormap of the best parameters $\beta$, $F_0$, $\sigma_0$ and $\alpha$ obtained by comparing each experiment with simulations. The color scale is normalized separately for each parameter  (with $\beta^{\rm max}=100$,  $F_0^{\rm max}=50$ and $\sigma_0^{\rm max}=400$ and $\alpha^{\rm max}=14.2$) Each column represents an experiment with a cell line associated with the color legend.
}
 \end{figure}

We find the best match between experiments and simulations (see supplementary methods)
and identify a set of four parameters  ($F_0$, $\sigma_0$, $\beta$, $\alpha$) that best describe each individual experiment. Representative examples of the good match between experiments and model are reported in Fig. \ref{fig:1} for the case of MCF-10A cells under different conditions. We have also checked that the simulations reproduce
the value of the correlation length with reasonable accuracy.
The collection of parameters obtained for all the experiments is summarized in Fig. \ref{fig:2}b 
as normalized color maps. The map allows us to observe 
some general trends in the parameter space: in particular MCF10-A with RAB5A overexpression stand out for their higher value of self-propulsion ($F_0$) and mutual alignment ($\beta$)  with respect to the other cells. Furthermore, MCF-10A, both WT and RAB5A, display weaker  noise ($\sigma_0$) in comparison with the other cells. 

 \begin{figure*}[htb]\centering 
\includegraphics[width=15cm]{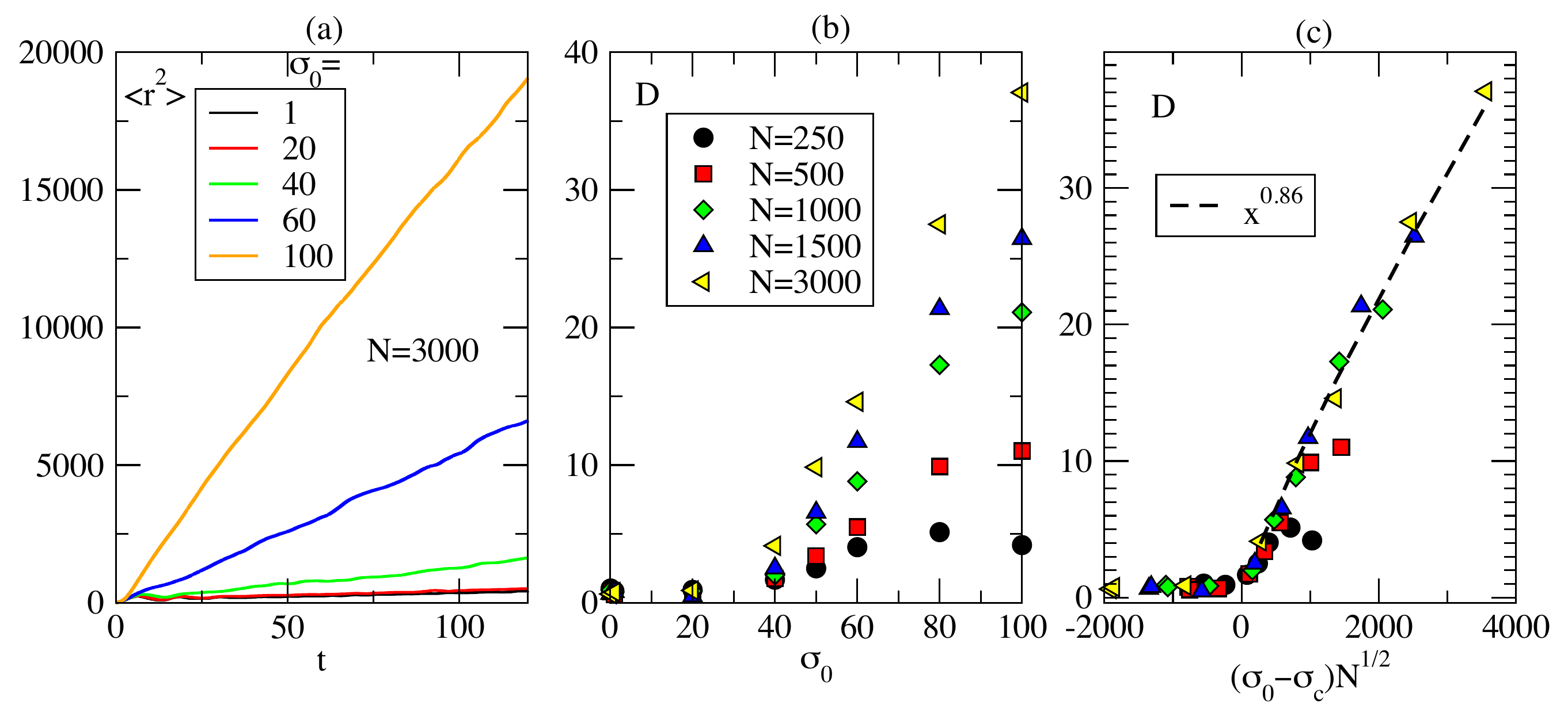}
 \caption{\label{fig:3} Characterization of the jamming-flowing transition in a
 simulated confluent layer.
 a) The time dependence of the mean-square displacement as a function of the noise amplitude 
 $\sigma_0$ b) The diffusion constant displays a system-size dependent behavior above a noise
 level $\sigma_c$ that can be used to identify the jamming point. c) The curves from panel b)
 can be collapsed assuming a finite-size scaling form.}
 \end{figure*}

\subsection{Finite size scaling at the jamming transition}
\begin{figure}[htb]\centering 
\includegraphics[width=\columnwidth]{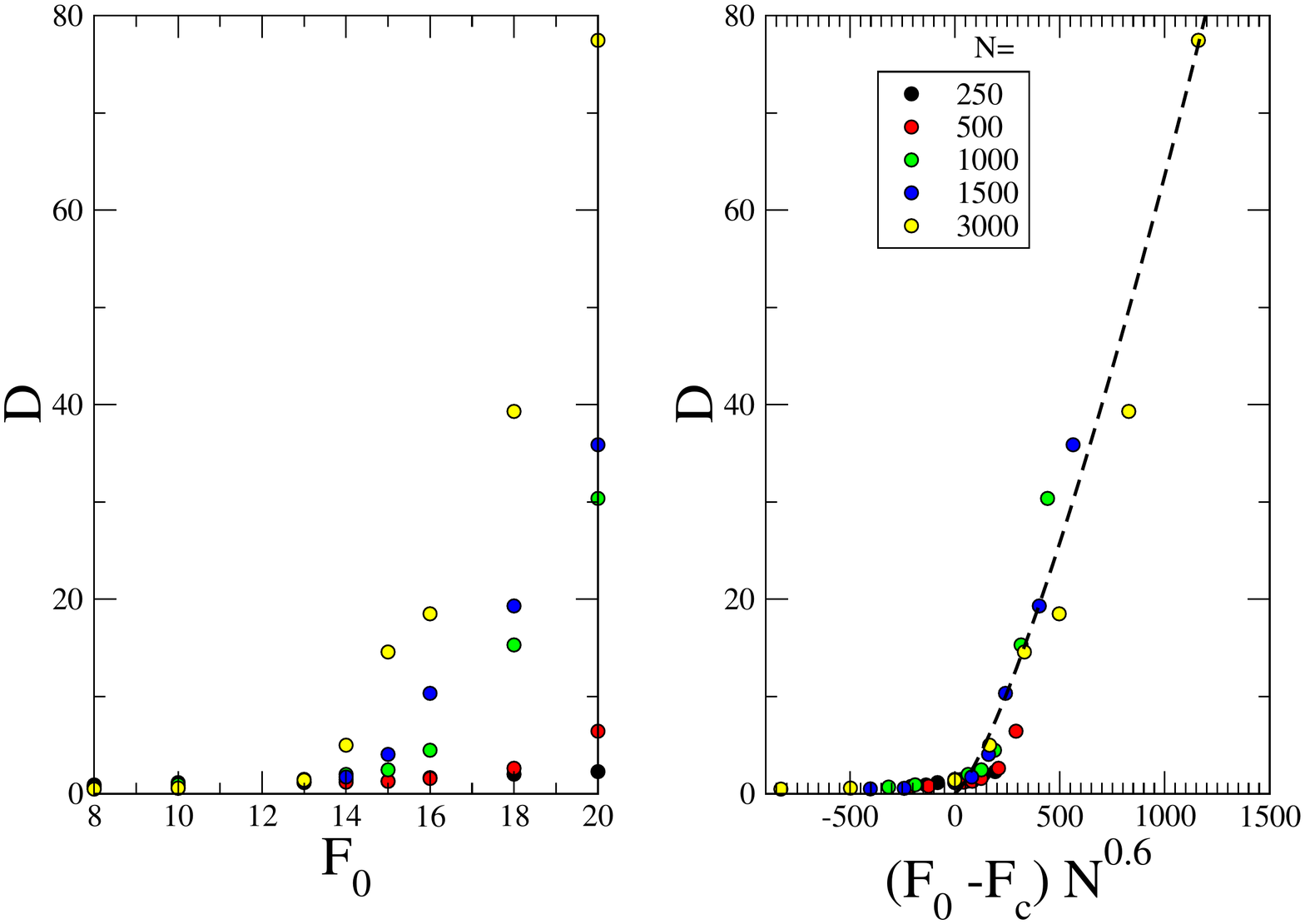}
 \caption{\label{fig:S3} a) The diffusion constant as a function of the self-propulsion force
 $F_0$
 displays a system-size dependent behavior above a noise
 level $F_c$ that can be used to identify the jamming point. b) The curves from panel a)
 can be collapsed assuming a finite-size scaling form.}
 \end{figure}

A simple inspection of the experimentally measured velocity distributions does not provides a clear understanding of the possible jamming or flowing behavior of the cell monolayers. Simulating the model, however, allows for a complete characterization of its dynamical behavior in term of a reduced set of parameters. Following recent work, we analyze the mean-square displacement (MSD) of individual cells in simulations of confluent layers and study how it changes as key parameters are varied. 
For instance, Fig. \ref{fig:3}a shows that the MSD is a linear function of time whose slope,
the effective diffusion constant $D$,  increases with noise parameter $\sigma_0$. A sharp increase
in $D$ is usually associated with a transition from a jamming to a flowing state. In our case,
Fig. \ref{fig:3}b shows that $D$ depends on $\sigma_0$ in a system-size dependent manner. This is generally expected for phase transitions in and out of equilibrium, where finite size scaling governs 
the behavior of the system near the critical point. We thus propose a finite-size scaling form
for the noise and size dependence of the diffusion constant 
\begin{equation}
D(\sigma_0,N) = f((\sigma_0-\sigma_c) N^{1/\nu}) 
\label{eq:collapse}
\end{equation}
where the scaling function $f(x) \propto x^\gamma$ for $x>0$ and $f(x)=\textrm{const.}$ for 
$x<0$. The data collapse reported in Fig. \ref{fig:3}c according to Eq. \ref{eq:collapse}
indicates that $1/\nu=0.5$ and $\gamma=0.85$. A similar finite-size scaling collapse 
can be done in dependence on $F_0$: $D(F_0,N) = g((F_0-F_c) N^{1/\nu'})$, with 
$g(x) \propto x^{\gamma'}$ for $x>0$ (see Fig. \ref{fig:S3}), yielding $1/{\nu'}=0.5$ and 
$\gamma'=1.3$. 

\subsection{Wound healing induces unjamming}
\begin{figure}[htb]\centering 
\includegraphics[width=\columnwidth]{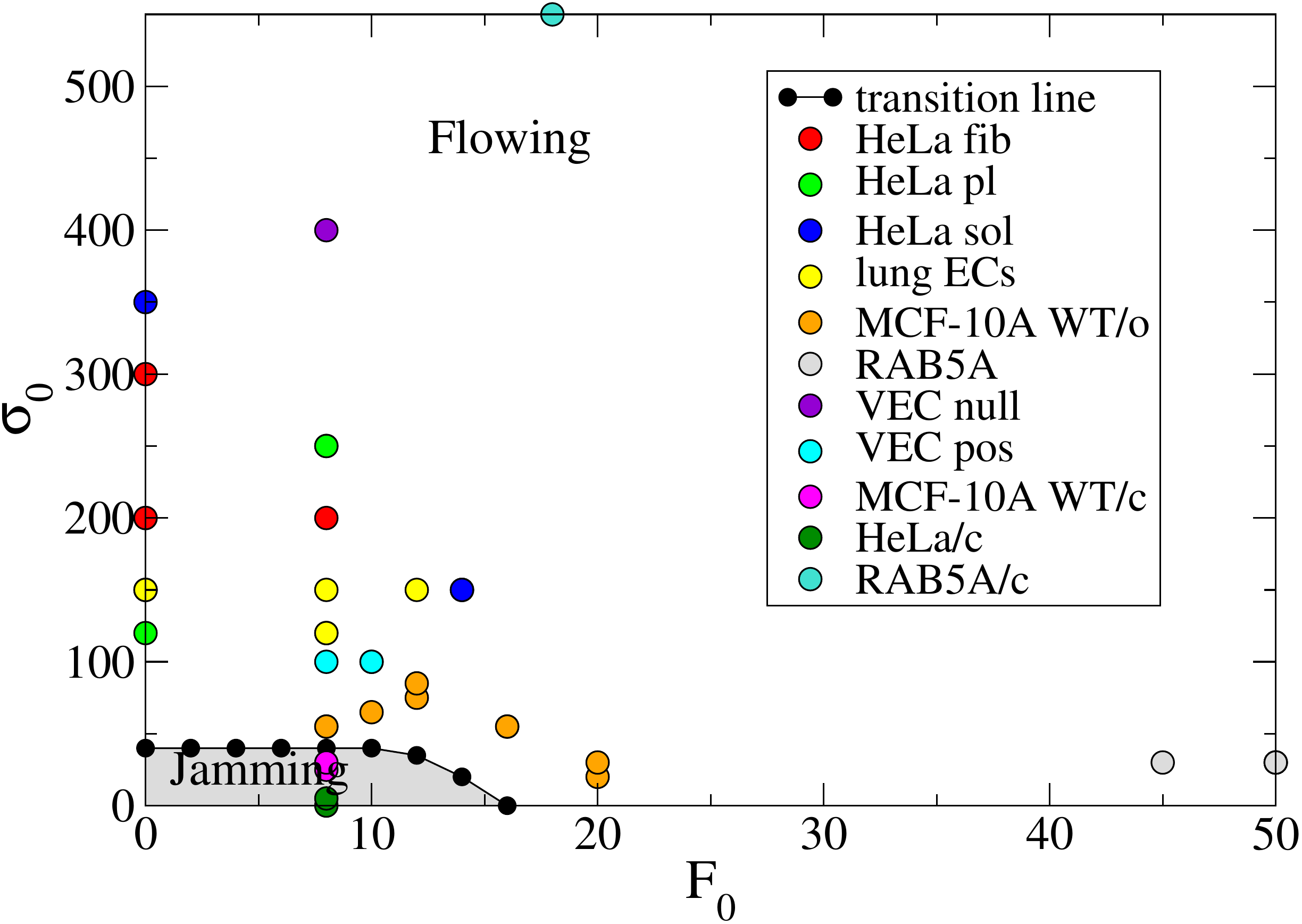}
 \caption{\label{fig:4} The jamming phase diagram. A projection of the jamming phase diagram
 on the ($\sigma_0$,$F_0$) plane, for $\beta=10$ and $\alpha=1.4$. For each experiment, we report 
 on the phase diagram the best estimate for $\sigma_0$ and $F_0$. Experiments performed in confluent
 conditions are all in the jamming phase, those in wound healing conditions are mostly in 
 the flowing phase.
 }
 \end{figure}

Collecting together all the simulations performed, we construct
a phase diagram distinguishing the jammed from the flowing state as a function of
$F_0$ and $\sigma_0$, as reported in Fig. \ref{fig:4}. Here for practical reasons, we have
concentrated on fixed values for $\beta=10$ and $\alpha=1.4$ that correspond 
to a large set of experimental data. It is possible in principle to reconstruct
the whole 4-dimensional phase diagram but this is extremely demanding computationally.
 
We are now in position to reach the goal posed at the beginning of this paper and identify the phase for each experimental sample. To this end, we report in Fig. \ref{fig:4} all the parameter values ($F_0$,$\sigma_0$) extracted from the experimental measurement (see Table S1 for the
parameter list). The general trend we observe from the figure is that all the experiments performed
in confluent layers fall into the jammed phase, while almost all of the wound healing experiments 
are in the flowing phase. Notice that here we are projecting all the data into the 
($\beta=10$, $\alpha=1.42$) plane so discrepancies could also arise due to the fact that experimental
data are described by different values of $\beta$ and $\alpha$. 

Taken together our results suggest that when a jammed confluent layer is perturbed by wound healing,
providing additional space for the cells, the system undergoes a phase transition to the flowing state. This is reflected by the fact that key parameters describing self-propulsion, noise or alignment change their values, implying that the monolayer responds in an active way to the
change of boundary conditions. We checked that this process is not an artifact of the model or the fitting by considering wound healing before and after closure. Results show that it is possible
to fit the velocity distributions before and after healing with the same parameters. It is only 
some time after closure that the cell layer slowly falls back into the jammed state.

\section{Discussion}

Soft and glassy materials are known to undergo a jamming transition, characterized by
limited mobility and slow relaxation upon fine tuning of a set
of physical control parameters including density, temperature and shear stress \cite{Liu2010}. 
Dense cellular assemblies such as epithelial tissues and cancer are recently emerging as
paradigmatic examples of jamming in active matter \cite{Angelini2010,Garcia2015,Park2015}, a well studied
non-equilibrium state of matter where an internal force drives each of the elementary 
units of the system. Examples of active matter range from artificial self-propelled colloidal
particles, to bacteria, epithelial and cancer cells  or even 
large groups of animals such as fish schools or birds flocks. A key issue for active matter
is to understand what are the main physical determinants for jamming, which would play
a role analogous to stress, density and temperature in conventional matter.

Current understanding of the jamming transition in cellular systems comes mostly from
vertex models \cite{Bi2015} and SVP models \cite{Li2014,Bi2016}, which allow to clarify
the role of many key biophysical parameters, such as cell anisotropies and self-propulsion
strength. These models, however, are defined for systems with periodic boundary conditions
and are therefore not appropriate to study the response of jammed epithelia to a wound, 
a well studied protocol to understand and quantify collective cell migration in epithelial and cancer cells 
\cite{Tambe2011,Brugues2014,Haeger2014,Lange2013,koch2012,Chepizhko2016,Poujade2007,Sepulveda2013,Vedula2013}.
Active particle models have proven very effective to quantitatively describe wound healing
\cite{Sepulveda2013,Chepizhko2016}, but are usually not employed to study the jamming
transition. Hence, the relation between wound healing and the jamming transition has not been explored. In the paper, we fill this gap by a combination of experiments and
numerical simulations.

We analyze a large set of time-lapse experiments of cellular assemblies 
performed for different cell lines both in confluent conditions or after wound healing. In all these case, we compute the velocity distributions by PIV
and study how the distribution changes in presence of a wound. We start our
analysis from the MCF-10A cell line that was recently shown to undergo a transition
from a jammed to a flowing state by over-expression of RAB5A, a master regulator of
endocytosis \cite{Malinverno2017}. Here we compare this transition with the 
similar fluidization induced by a wound. To obtain a quantitative characterization of the transition, we compare the experimental results with numerical simulations of an active
particle model. In this way, we obtain the model parameters that best describe
the experimentally measured velocity distributions. This parameters describe
the self-propulsion force, the alignment interactions, intracellular adhesion, damping and noise and allow for a physical characterization of different
cell lines and experimental conditions. 

Our first observation is that the internal parameters typically change when a wound is present. This is a typical feature of living matter and not shared by ordinary 
inanimate matter where boundary conditions are not expected to influence the internal parameters of the system. Thus a jammed solid could possibly slowly invade a new open space and unjamming would eventually only occur because of a reduction of the density. In our case, however,  the change of velocity distribution is instead compatible with a rapid change of internal parameters such as self-propulsion or alignment, way before the 
cell density is reduced. Once the parameters associated to each experiments have been determined, we are able to place them on a phase diagram reconstructed
with the model. To this end, we perform numerical simulations of the active particle model under
closed boundary conditions. The jamming phase can be identified considering the particle 
 diffusion, quantified by the mean-square displacement \cite{Bi2016}. We observe that
 the effective diffusion constant displays system size dependent effect in parameter space,
 allowing us to identify the transition point. As expected for second-order phase transitions, the 
 curve describing the variation of the diffusion constant with control parameters and  
 system size obeys finite size scaling. We can thus obtain scaling exponents by 
 performing data collapse. 

In our analysis of wound healing experiments, we have recorded parameters in steady-state conditions,
disregarding the initial transient phase where the horizontal velocity builds up. Previous studies have
shown that collective migration is associated with the propagation of waves of coordinated reorientation
starting at the front \cite{Zaritsky2014,Zaritsky2015}. This is in line with our observation of
a change of parameters induced by the boundary and suggests that changes first occur at the wound site
and then rapidly propagate inside the layer. The phenomenon of collective cell velocity reorientation in the
direction of the wound has been explained considering cell guidance through intracellular force propagation
in combination with the idea of plithotaxis, defined as the tendency for cells to move along the
axis of maximal normal stress \cite{Trepat2011}. The presence of a wound suddenly decreases the local
stress on the boundary cells, inducing internal changes that are then transmitted to the 
bulk cells, leading to the emergent unjamming of the entire monolayer.

\appendix
\section{Methods \label{sec:methods}}

\subsection*{Cell culture}
HeLa cell line (ATCC CCL-2) is cultured as discussed in Ref. \cite{Chepizhko2016}.  Cells are seeded in bovine soluble collagen, or fibrillar bovine collagen or without collagen-coated dishes and growth up to reach confluence overnight (see \cite{Chepizhko2016})
Endothelial cells derived from embryonic stem cells with homozygous null
mutation of the VE-cadherin gene (VEC null) \cite{balconi2000}. The
wild type form of VE-cadherin was introduced in these cells (VEC positive) as
described in detail in \cite{Lampugnani2002}.  
Endothelial cells isolated from lungs of wild type adult mice and 
cultured as previously described \cite{giampietro2015,Chepizhko2016}.  
The MCF-10A cell line is a non-tumorigenic human mammary epithelial cell
line cultured as discussed in \cite{Malinverno2017}.

\subsection*{Wound healing assay}
For the migration assay, a wound is introduced in the central area of the confluent cell sheet by using a pipette tip and the migration followed by time-lapse imaging. Hela cells were stained with 10 $\mu$M Cell tracker green CMFDA (Molecular Probes) in serum-free medium for 30 minutes and then the complete medium was replaced. Mouse endothelial cell monolayers were wounded after an overnight starving, washed with PBS, and incubated at 37$^\circ$C in starving medium. MCF-10A cell monolayers were wounded after an O/N doxycicline induction, washed with PBS, and incubated at 37$^\circ$C in fresh media+doxycicline.

\subsection*{Time lapse imaging}
For Hela cells, time-lapse multifield experiments were performed using an automated inverted Zeiss Axiovert S100 TV2 microscope (Carl Zeiss Microimaging Inc., Thornwood, NY) with a chilled Hamamatsu  CCD camera OrcaII-ER. Displacements of the sample and the image acquisition  are computer-controlled using Oko-Vision software (from Oko-lab).  This Microscope was equipped with a cage incubator designed to maintain all the required environmental conditions for cell culture all around the microscopy workstation, thus enabling to carry out prolonged observations on biological specimens. Cell Tracker and Phase contrast images were acquired with an A-Plan 10x (NA 0.25) objective; the typical delay between two successive images of the same field was set to 10 minutes for 12 hours. 
For mouse endothelial cells and MCF-10A cells, time-lapse imaging of cell migration was performed on an inverted microscope (Eclipse TE2000-E; Nikon) equipped with an incubation chamber (OKOLab) maintained at 37$^\circ$C in an atmosphere of 5\% CO2. Movies were acquired with a Cascade II 512 (Photometrics) charge-coupled device (CCD) camera controlled by MetaMorph Software (Universal Imaging) using a 4X or 10× magnification objective lens (Plan Fluor 10×, NA 0.30). Images were acquired every 2 or 5 min over a 24h period. ters. See Refs. \cite{Chepizhko2016,Malinverno2017} for more details.

\subsection*{Particle image velocimetry (PIV)}
The measurements of the velocity field were done using PIVlab app for Matlab \cite{PIV0, PIV1}. The method is based on the comparison of the intensity fields of two consequent photographs of cells. The difference in the intensity is converted into velocity field measured in $px/frame$ and then converted to $\mu m/h$ \cite{Chepizhko2016}.


\subsection*{Fit of the experimental data}
To find the values of simulation parameters that fit experimental data, extensive numerical simulations were performed to cover the parameter space. 
The main parameters to vary were $\beta$, $\sigma_0$, $F_0$, and, to less extent, $\alpha$. 
Then the fitting was done for each experimental pair of values $\langle v_x \rangle_{exp}$ and $\Sigma_{exp}$,  through the variable $M=
((\xi_{exp}-\xi_{sim})/\xi_{exp})^2+ \int dv (P_{exp}(v)- P_{sim}(v))^2/\Sigma_{exp}$. The variable $M$ 
was computed for a large number of simulations spanning the parameter space and its minimum was selected as the best fit.


%

\end{document}